# Superconductivity and Charge Density Wave of $CuIr_2Te_4$ by Iodine Doping*

Mebrouka Boubeche[1†], Jia Yu(于佳)[2†], Chushan Li(李楚善)[2], Huichao Wang(王慧超)[2], Lingyong Zeng(曾令勇)[1], Yiyi He(何溢懿)[1], Xiaopeng Wang(王晓鹏)[1], Wanzhen Su(苏婉珍)[1], Meng Wang(王猛)[2], Dao-Xin Yao(姚道新)[2], Zhijun Wang(王志俊)[3,4] and Huixia Luo(罗惠霞)[1**]

[1]School of Materials Science and Engineering, State Key Laboratory of Optoelectronic Materials and Technologies, Key Lab of Polymer Composite & Functional Materials, Sun Yat-Sen University, No. 135, Xingang Xi Road, Guangzhou, 510275, P. R. China
[2]School of Physics, Sun Yat-Sen University, No. 135, Xingang Xi Road, Guangzhou, 510275, P. R. China
[3]Beijing National Laboratory for Condensed Matter Physics, and Institute of Physics, Chinese Academy of Sciences, Beijing 100190, China
[4]University of Chinese Academy of Sciences, Beijing 100049, China
† These authors contributed equally to this work; email: boubeche@mail.sysu.edu.cn; yujia7@mail.sysu.edu.cn
**Corresponding author Email: luohx7@mail.sysu.edu.cn; Tel: (+0086) 13802768250

*H.X. Luo was supported by the National Natural Science Foundation of China (Grants No. 11922415), Guangdong Basic and Applied Basic Research Foundation (2019A1515011718), the Fundamental Research Funds for the Central Universities (19lgzd03), Key Research & Development Program of Guangdong Province, China (2019B110209003), and the Pearl River Scholarship Program of Guangdong Province Universities and Colleges (20191001). H.C. Wang was supported by the National Natural Science Foundation of China (Grant Nos. 12004441), the Hundreds of Talents program of Sun Yat-Sen University and the Fundamental Research Funds for the Central Universities (No. 20lgpy165). D.X. Yao was supported by the National Natural Science Foundation of China (Grants No.11974432), NKRDPC-2017YFA0206203, NKRDPC-2018YFA0306001. M. Wang was supported by the National Nature Science Foundation of China (11904414), Natural Science Foundation of Guangdong (2018A030313055), and National Key Research and Development Program of China (2019YFA0705700).

Here we report a systematic investigation on the evolution of the structural and physical properties, including the charge density wave (CDW) and superconductivity of the polycrystalline $CuIr_2Te_{4-x}I_x$ for $0.0 \leq x \leq 1.0$. X-ray diffraction results indicate that both of $a$ and $c$ lattice parameters increase linearly when $0.0 \leq x \leq 1.0$. The resistivity measurements indicate that the CDW is destabilized with slight $x$ but reappears at $x \geq 0.9$ with very high $T_{CDW}$. Meanwhile, the superconducting transition temperature ($T_c$) enhances as $x$ raises and reaches a maximum value of around 2.95 K for the optimal composition $CuIr_2Te_{3.9}I_{0.1}$ followed by a slight decrease with higher iodine doping content. The specific heat jump ($\Delta C/\gamma T_c$) for the optimal composition $CuIr_2Te_{3.9}I_{0.1}$ is approximately 1.46, which is close to the Bardeen-Cooper-Schrieffer (BCS) value which is 1.43, indicating it is a bulk superconductor. The results of thermodynamic heat capacity measurements under different magnetic fields ($C_p(T, H)$), magnetization $M(T, H)$ and magneto-transport $\rho(T, H)$ measurements further suggest that $CuIr_2Te_{4-x}I_x$ bulks are type-II superconductors. Finally, an electronic phase diagram for this $CuIr_2Te_{4-x}I_x$ system has been constructed. The present study provides a suitable material platform for further investigation of the interplay of the CDW and superconductivity.

Layered transition metal dichalcogenides (TMDCs) as low dimensional solids have long been interested in multitudinous research fields such as energy, sensing, electronics and environmental, due to their tunable band gaps and natural abundancy. [1-4] Especially, the interplay between charge density wave (CDW) and superconductivity (SC) in TMDC materials has been one of the crucial research topics in condensed matter physics. [5-8] Importantly, the superconducting and CDW transition temperatures can be readily tuned by chemical doping [9-12] or applying high pressure [13-15] or field gating [16-18], resulting in dome-shape phase diagrams. [13,19] This phenomenon can be easily reminiscent of the classic cuprates and iron-based high-temperature superconductors, in which the antiferromagnetic states can be turned into SC by controlling system parameters and eventually form dome-shaped phase diagrams. [20-21] Naturally, we notice that CDWs in TMDC materials perceived a distinct similarity to those antiferromagnetic orders in high-temperature cuprates and iron-based superconductors. Therefore, systematic research on TMDCs may provide further insights into the relevance of non-Bardeen-Cooper-Schrieffer (BCS) behaviors in unconventional superconductors. [22-23]

Recently, the layered $IrTe_2$, which undergoes a first-order structural phase CDW-like transition at around 250 K, has attracted considerable interest. A few examples showed that the superconductivity emerges at around 3 K upon suppression of the CDW-like phase transition by Cu or Pd intercalation, [24-25] or with partial metallic substitution of Pd, Pt or Rh in the Ir-site, [25-29] accompanied by the monoclinic phase of $IrTe_2$ tuning into the trigonal phase. For example, M. Kamitani et al. proposed that Cu intercalation into the pristine $IrTe_2$ ($Cu_xIrTe_2$, $0 \leq x \leq 0.12$) acts as electron doping. The structural/electronic phase transition starts to be suppressed by the Cu intercalation of $x = 0.03$. It almost disappears for $x = 0.05$, where the trigonal phase is stabilized down to the lowest temperature, and superconductivity with $T_c = 2.8$ K is induced. [24]

More recently, we have successfully synthesized a quasi-two-dimensional (2D) $Cu_{0.5}IrTe_2$ (so far indexed as $CuIr_2Te_4$) with defect NiAs structure of trigonal symmetry (space group *P3-m1*) as shown in Figure 1(a), embodying 2D $IrTe_2$ layers and intercalated by Cu between the layers. This copper telluride chalcogenide exhibits the coexistence of the superconductivity ($T_c = 2.5$ K) and CDW-like transition ($T_{CDW} = 250$ K), [30] where the first principle calculations imply that the states near the Fermi energy mainly come from Te $p$ and Ir $d$ orbitals. Afterward, we studied the influence of different dopants with $3s$ to $5d$ (e.g. Al, Ti, Zn, Ru, Rh) in the Cu-site or the Ir-site of $CuIr_2Te_4$ [31-32] on the structural and superconducting properties and found that they behave quite differently, which may have broad implications in the search for new superconductors and quantum states. Nevertheless, up to now, a lack of studies on the effect of halogen group dopants on TMDCs has been reported. Yet, the halogen elements (e.g. F, Br and Cl) have been found to increase the $T_c$ in high-$T_c$ iron-based superconductors successfully (e.g. $La(O_{1-x}F_x)FeAs$) [33-34] and misfit phase superconductor $Pb_{1.12}Ta(Se_{0.92}X_{0.08})_{3.1}$ ($X$ = Cl and Br). [35] Besides, a few I-based chalcogenides such as $(TaSe_4)_2I$, [15] $Bi_2TeI$, [36] and $Bi_4I_4$, [37] became hot frontier materials due to their unique topological quantum properties. Therefore, whether superconductivity and CDW in $CuIr_2Te_4$ can be tuned or new quantum states can be triggered by 5$p$ iodine doping becomes a challenging question.

In this study, we focus on the effect of iodine substitution for Te-site in CuIr$_2$Te$_4$ on the superconductivity and charge density wave through X-ray diffraction (XRD) analysis, electronic transport, magnetic susceptibility and specific heat measurements. A detailed electronic phase diagram for CuIr$_2$Te$_{4-x}$I$_x$ (0.0 ≤ $x$ ≤ 1.0) has been proposed. Unexpectedly, the signature of CDW is absent even if slight iodine $x$ is substituted for Te, but reemerges at $x$ ≥ 0.9 with very higher $T_{CDW}$ than that of the host compound. Superconducting transition temperature ($T_c$) enhances slightly as $x$ raises and reaches the maximum value of 2.95 K for the optimal composition CuIr$_2$Te$_{1.9}$I$_{0.1}$, and then decreases slightly with higher iodine doping content. The normalized specific heat jump ($\Delta C/\gamma T_c$) for the optimal composition CuIr$_2$Te$_{3.9}$I$_{0.1}$ is approximately 1.46, which is close to the Bardeen-Cooper-Schrieffer (BCS) value of 1.43, indicating it is a bulk superconductor. The results of thermodynamic heat capacity measurements under different magnetic fields ($C_p(T, H)$), magnetization $M(T, H)$ and magneto-transport $\rho(T, H)$ measurements further suggest that our CuIr$_2$Te$_{4-x}$I$_x$ are type-II superconductors. Finally, the electronic phase diagram for this CuIr$_2$Te$_{4-x}$I$_x$ system has been proposed, which may be helpful for further investigation of the interplay of the CDW and superconductivity.

Standard solid-state reactions were employed to synthesize polycrystalline samples of CuIr$_2$Te$_{4-x}$I$_x$ (0.0 ≤ $x$ ≤ 1.0). The mixture of high-purity powders of Cu (99.9%), Ir (99.9%), Te (99.999%), CuI (99.9%) and I$_2$ (99.9%) in the appropriate stoichiometric ratios were heated in sealed evacuated silica glass tubes by the rate 1 °C/min to 850 °C for 120 h. Afterward, the as-prepared powders were reground, pelletized and sintered again in the same conditions for 240 h with intermediate grinding.

The crystallographic structure and phase purity of our samples were determined by powder X-ray diffraction (PXRD) MiniFlex, Rigaku with Cu $K\alpha$1 radiation. The lattice parameters profile fits were obtained after Rietveld refinement using Thompson-Cox-Hastings pseudo-Voigt peak shapes by the FULLPROF suite software. [38] Measurements of the temperature-dependence of electrical resistivity (four-point method), specific heat, and magnetic susceptibility were carried out using a DynaCool Quantum Design Physical Property Measurement System (PPMS). $T_c$ is estimated conservatively: for susceptibility, $T_c$ was taken as the intersection of the extrapolations of the abrupt slope of the susceptibility in the superconducting transition region and the normal state susceptibility; for resistivities, the midpoint of the resistivity $\rho(T)$ transitions was taken, and, for the specific heat data, the critical temperatures obtained from the equal area construction method. The elements distributions and ratios of samples were analyzed using scanning electronic microscope (SEM) EM-30AX PLUS from Kurashiki Kako Co. Ltd, Japan, equipped with an Energy Dispersive X-ray Spectroscopy (EDXS) detector.

PXRD examination was performed to inspect the structure and phase purity of synthesized polycrystalline CuIr$_2$Te$_{4-x}$I$_x$ (0.0 ≤ $x$ ≤ 1.0). Figure 1 (b) and Table 1 show the detailed Rietveld refinement results of the representative CuIr$_2$Te$_{3.9}$I$_{0.1}$ sample, which illustrates that the reflection peaks correspond to the trigonal phase of CuIr$_2$Te$_{4-x}$I$_x$ with space group $P3-m$1, and there is some unreacted Ir as an impurity. From Figure 1 (c), it can be seen that the obtained powder CuIr$_2$Te$_{4-x}$I$_x$ (0.0 ≤ x 1.0) samples mainly consist of the trigonal phase. However, the unreacted Ir impurities are present in all the samples, and the Ir peak becomes stronger for higher doping concentration $x$. The detailed refinements for the polycrystalline samples are represented in Figure S1. The molar percentages of impurity (Ir) for $CuIr_2Te_{4-x}I_x$ (0 ≤ x ≤ 1) obtained from the Rietveld refinements is found to be less than 3.8 % but is getting bigger with higher doping concentration, suggesting that the higher iodine doping concentration increases the disorder in the crystalline structure. The amplification of the (002) peak in Figure 1 (c) shows an apparent left shift with the increasing $x$. Figure 1 (d) shows the doping amount dependence of the lattice parameters. Obviously, the lattice constants $a$ and $c$ for the investigated system increase linearly as doping contents increase (0.0 ≤ x ≤ 1.0), which adapts the Vegard's law. [40] Further, we use EDXS to characterize the investigated compounds and find the experimental ratios of Cu:Ir:Te:I for $CuIr_2Te_{4-x}I_x$ samples with 0 ≤ x ≤ 1 are close to the calculated ratios (see Figure S2 and Table S1).

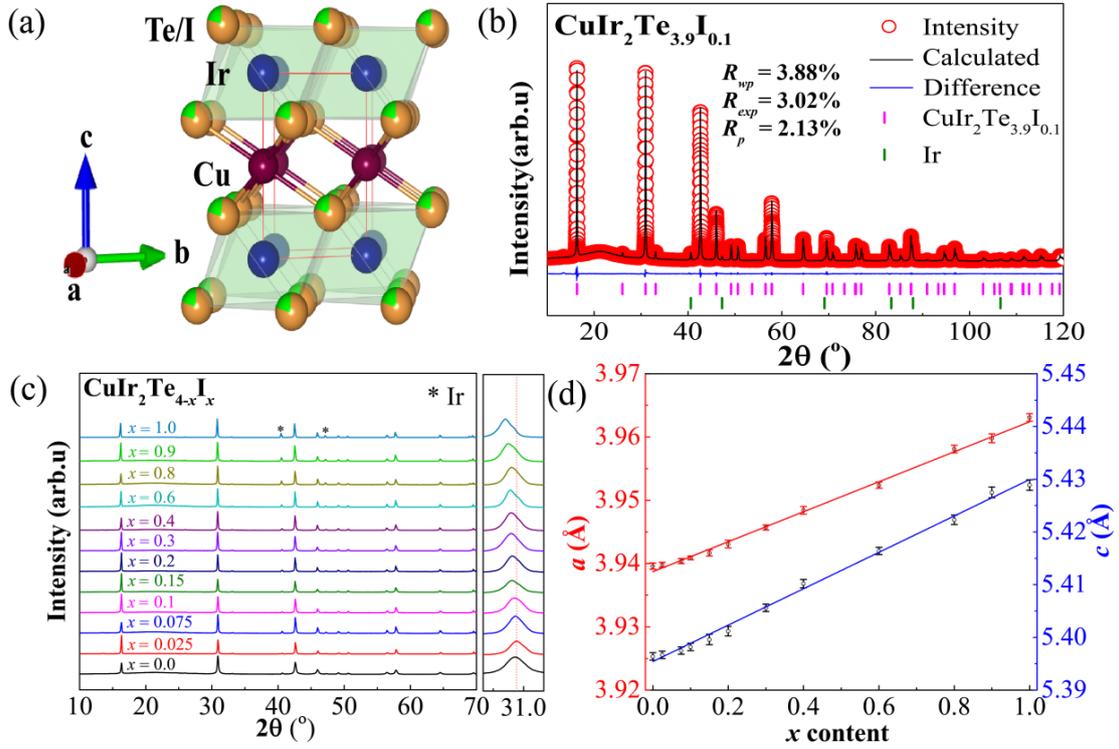

**Fig. 1.** Structural characterization and analysis of $CuIr_2Te_{4-x}I_x$. **(a)** The crystal structures of $CuIr_2Te_{4-x}I_x$. **(b)** The powder XRD pattern with Rietveld refinement for $CuIr_2Te_{3.9}I_{0.1}$. **(c)** Powder X-ray diffraction patterns for $CuIr_2Te_{4-x}I_x$ (0.0 ≤ x ≤ 1.0). **(d)** The variation of in-plane lattice parameters, $a$ and $c$, with $x$ content.

**Table 1**. Rietveld refinement structural parameters of $CuIr_2Te_{3.9}I_{0.1}$. Space group *P3-m1* (no. 164), a = b = 3.9409(2) Å and c = 5.3982 (3) Å, $R_p$ = 3.88%, and $R_{wp}$ = 3.02%, $R_{exp}$ = 2.13%.

| Label | x | y | z | site | Occupancy |
|---|---|---|---|---|---|
| Cu | 0 | 0 | 0.5 | 2b | 0.5 |
| Ir | 0 | 0 | 0 | 1a | 1 |
| Te | 0.33333 | 0.66667 | 0.25292(3) | 2b | 0.973(3) |
| I | 0.33330 | 0.66667 | 0.25291(3) | 2d | 0.024(2) |

Next, we carried out the electronic transport $\rho(T)$ and magnetic susceptibility $M(T)$ measurements on $CuIr_2Te_{4-x}I_x$ to determine the superconducting transitions. The temperature dependence of the real electrical resistivity $\rho(T)$ and the normalized resistivity ($\rho/\rho_{300K}$) curves are displayed in Figure 2 (a) and (b), respectively. The resistivity data show a metallic behavior for all the samples when the temperature is higher than 3 K. The resistivity for $CuIr_2Te_{4-x}I_x$ ($0.0 \leq x \leq 1.0$) decreases sharply to zero at lower temperatures, implying an emergence of superconductivity (Figure 2 (c)). However, no superconducting transition can be observed in the samples with high doping ($0.9 \leq x \leq 1.0$) above 2.1 K. The transition temperature ($T_c$) slightly increases and reaches a maximum value of 2.95 K at $x = 0.1$, which is somewhat higher than the $T_c$ the pristine $CuIr_2Te_4$ (2.5 K) [30] and $T_c$s obtained by the optimal Ru (2.79 K), [31] Zn (2.82 K), Ti (2.84 K) and Al (2.75 K) substitution, [32] followed by a slight decrease of $T_c$ as $x$ increases. The change of $T_c$s is also clearly seen in the zero Field cooling (ZFC) magnetic susceptibility data at low temperatures, as shown in Figure 2 (d). We can estimate the superconducting volume fraction for all samples > 80 %, revealing the high purity of our $CuIr_2Te_{4-x}I_x$ samples. The ZFC magnetic susceptibility combining with the following heat capacity measurements imply bulk superconductivity. Unexpectedly, the signature of the CDW-like transition disappears for the samples with even a small doping amount, while the CDW-like transition reappears at high doping content ($0.9 \leq x \leq 1$), as depicted in Figure 2 (a) and (b). A hump-like anomaly in the $\rho(T)$ curve develops starting from $x = 0.9$ within the temperature interval 240-180 K in the higher doping matrices $CuIr_2Te_{4-x}I_x$ ($0.9 \leq x \leq 1.0$) (see Figure 2b). The inset of Figure 2 (b) presents the temperature derivative of resistivity ($d\rho/dT$) $vs.$ $T$ curve at high temperatures measured from a cooling process around the CDW-like humps, which is further used to estimate the CDW-like transition. In fact, a similar behavior has also been found in an early report in Tl-intermediate $Nb_3Te_4$ single crystals [40], attributed to disorder in the quasi 1-D Nb chains. Besides, $3d$ transition metal intercalated-$M_x TiSe_2$ (M = $3d$ transition metals, such as Mn, Cr) compounds also go through analogous phenomena, in which the first suppression and then reemergence of the CDW state with higher $3d$ metals intercalation concentration is probably ascribed to the degree of deformation of Se-Ti-Se sandwiches [41,42]. The reproduced CDW has also been found in the 1T-$TaS_{2-x}Se_x$ single crystals for heavy Se content [43]. Another prominent case is 2H-$TaSe_{2-x}S_x$ ($0 \leq x \leq 2$) [44] in which the dome-like superconducting phase diagram is large and superconductivity is enhanced with a crystallographic disorder, accompanied with the occurrence of CDW at two ends, where the disorder plays a crucial role in the evolution of CDW and SC. Consequently, by this comparison, the possible reappearance of the CDW-like transition in our $CuIr_2Te_{4-x}I_x$ ($0.9 \leq x \leq 1$) bulks is likely because of disordering effects created by iodine doping, which may be related to the possible occupied ordering between I and Te ions. However, the origin of the reappearance of the CDW-like transition is still an open question and requires further examination through (low-temperature) high resolution transmission electron microscopy (HRTEM), low-temperature powder neutron diffraction, X-ray scatter techniques, Raman and so on.

Besides, we have further conducted Hall measurements for three $CuIr_2Te_{4-x}I_x$ samples with $x = 0$, $x = 0.2$ and $x = 0.8$. The results for two representative samples are shown as Figure S3. The positive Hall coefficient reveals hole carrier type in our $CuIr_2Te_{4-x}I_x$ crystals before and after I-doping at low temperatures. For the undoped $CuIr_2Te_4$, the hole concentration is about $1.6 \times 10^{22}$ $cm^{-3}$ at 5 K. The doped $CuIr_2Te_{4-x}I_x$ crystals show higher hole carrier density of about $3.1\times10^{22}$ $cm^{-3}$ and $3.7\times10^{22}$ $cm^{-3}$ for $x = 0.2$ and $x = 0.8$ at 5 K, respectively. In addition, the Hall traces at different selected temperatures exhibiting close Hall coefficient indicate that the carrier density does not change much from 5 K to room temperature for $CuIr_2Te_4$ and $CuIr_2Te_{3.8}I_{0.2}$.

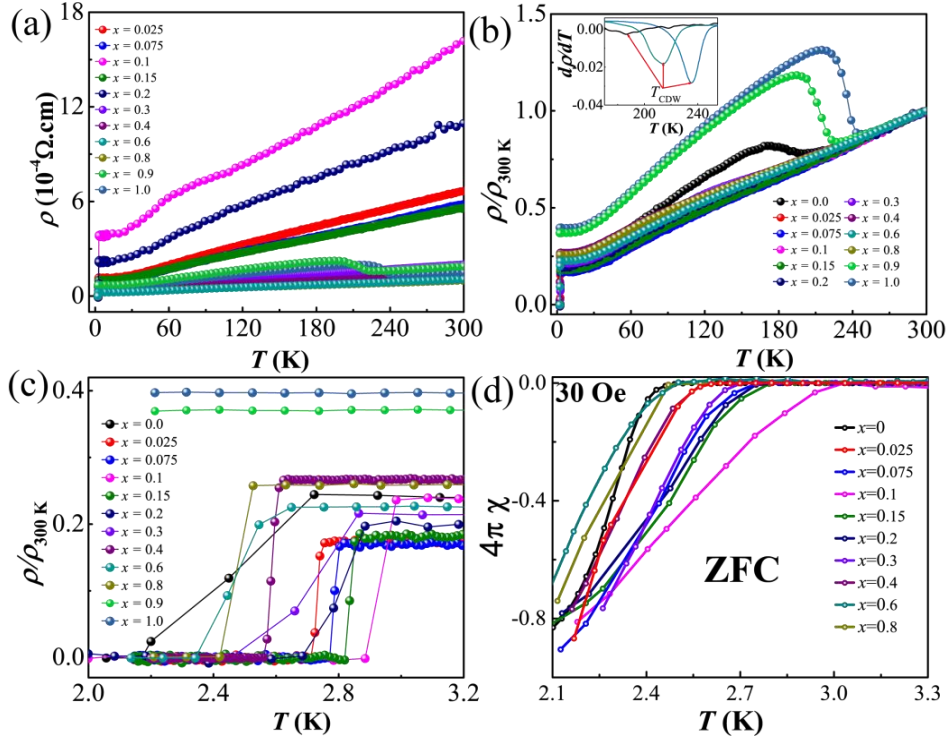

**Fig. 2.** Transport characterization of the normal states and superconducting transitions for $CuIr_2Te_{4-x}I_x$. **(a)** The temperature dependence of the resistivity for polycrystalline $CuIr_2Te_{4-x}I_x$ ($0.0 \leq x \leq 1.0$). **(b)** The temperature dependence of the resistivity ratio ($\rho/\rho_{300K}$) for polycrystalline $CuIr_2Te_{4-x}I_x$. **(c)** The temperature dependence of the resistivity ratio ($\rho/\rho_{300K}$) for polycrystalline $CuIr_2Te_{4-x}I_x$ at low temperatures. **(d)** The zero field cooling (ZFC) magnetic susceptibilities for $CuIr_2Te_{4-x}I_x$ ($0.0 \leq x \leq 0.8$) at the superconducting transitions with applied dc fields 30 Oe.

The low-temperature heat capacity measurements were performed to confirm the bulk nature of the superconductivity of the optimal superconducting $CuIr_2Te_{3.9}I_{0.1}$ polycrystalline sample to obtain more superconducting parameters. The specific heats at temperatures above $T_c$ to 10 K are well described as a sum of a $\beta T^3$ phonon contribution ($C_{ph.}$) and the $\gamma T$ electronic contribution ($C_{el.}$) based on the equation: $C_p = \gamma T + \beta T^3$. The data in Figure 3 (a) shows a prominent peak at the superconducting transition temperature for the optimal superconducting $CuIr_2Te_{3.9}I_{0.1}$, which is suppressed by a magnetic field of 1 T. The $T_c$ is extracted using an equal area entropy construction (solid blue lines) to be 2.94 K, which is consistent with those observed in the resistivity and susceptibility measurements in Figure 2. The normalized specific heat jump value ($\Delta C/\gamma T_c$) is estimated to be 1.46, which is very close to that of the BCS weak-coupling limit value (1.43), confirming bulk superconductivity. Further, $\beta$ value is near 3.03 mJ $mol^{-1}$ $K^{-4}$ and $\gamma$ is about 12.97 mJ mol$^{-1}$ K$^{-2}$. Debye temperature ($\Theta_D$) is calculated by using the formula $\Theta_D = (12\pi^4 nR/5\beta)^{1/3}$ where $n = 7$ is the number of atoms per formula unit and $R$ is the gas constant. The obtained $\Theta_D$ value for the CuIr$_2$Te$_{3.9}$I$_{0.1}$ sample is 164.9 K. Hence, the electron-phonon coupling constant ($\lambda_{ep}$) can be obtained using $\Theta_D$ and $T_c$ values from the inverted McMillan formula: [45]

$$\lambda_{ep} = \frac{1.04 + \mu^* \ln(\Theta_D/1.45T_c)}{((1 - 1.62\mu^*)\ln(\Theta_D/1.45T_c) - 1.04)},$$ where the repulsive screened coulomb parameter $\mu^*$ is 0.15. The estimated value of $\lambda_{ep}$ is about 0.70. Using the values of $\gamma$ and $\lambda_{ep}$, the density of states, $N(E_F)$, at the Fermi level can be calculated from the formula $N(E_F) = 3\gamma/(\pi^2 k_B^2 (1 + \lambda_{ep}))$, which is 3.24 states/$eV$ f.u. This value is higher than that of the host CuIr$_2$Te$_4$ and Ru doped CuIr$_2$Te$_4$ (see Table 2). Figure 3 (b) displays the magnetic field dependence of specific heat measurement for the optimal superconducting CuIr$_2$Te$_{3.9}$I$_{0.1}$ specimen with applied magnetic fields of 0, 0.01, 0.02, 0.03 and 0.04 T, plotted as $C_p/T$ vs. $T^2$. As predicted, the specific heat peak of the superconducting transition shifts to lower temperatures by increasing the applied magnetic fields.

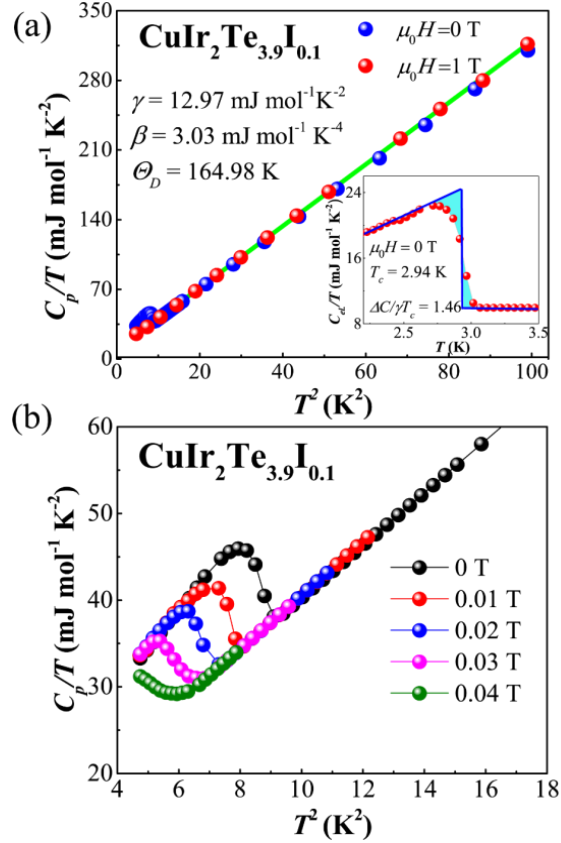

**Fig. 3.** Characterization of the superconducting transitions for CuIr$_2$Te$_{3.9}$I$_{0.1}$ through the measurement of the temperature-dependent specific heat. **(a)** $C_p(T)/T$ versus $T^2$ and the corresponding fitting with $C_p/T = \gamma + \beta T^2$. The inset shows ($C_{el.}/T$) vs $T$, where $C_{el.} = C_p - \beta T^3 = C_p - C_{ph.}$. **(b)** $C_p(T)/T$ vs $T^2$ for CuIr$_2$Te$_{3.9}$I$_{0.1}$ under various magnetic fields around the superconducting transition.

For the aim of estimating the lower critical field ($\mu_0 H_{c1}$), the field-dependent magnetization $M(H)$ measurements under different temperatures are carried out. Figure 4 shows the temperature dependence of $\mu_0H_{c1}$ for the selected CuIrTe$_{4-x}$I$_x$ ($x$ = 0.1, 0.15, 0.8) compounds. The insets at the upper right corners of Figure 4 show the detailed field-dependent magnetization $M(H)$ measurement

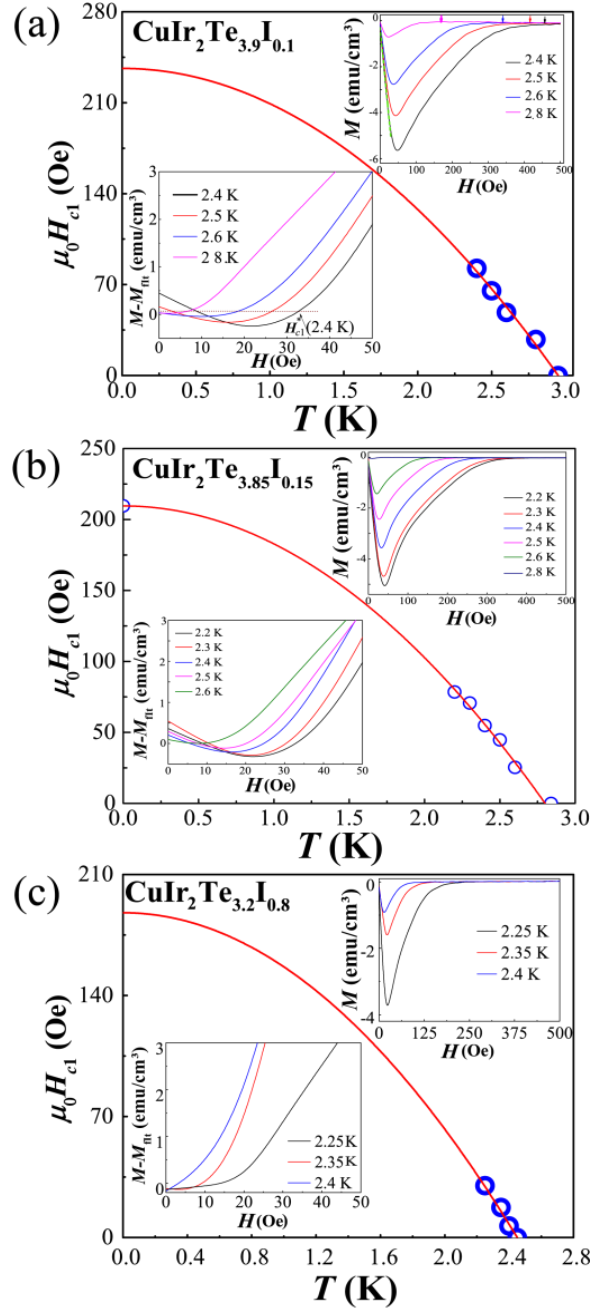

**Fig. 4 (a-c)** The lower critical fields for CuIr$_2$Te$_{3.9}$I$_{0.1}$, CuIr$_2$Te$_{3.85}$I$_{0.15}$ and CuIr$_2$Te$_{3.2}$I$_{0.8}$, respectively, with the fitting lines using the equation $\mu_0H_{c1}(T) = \mu_0H_{c1}(0)(1-(T/T_c)^2)$. The top insets show the magnetic susceptibilities $M(H)$ curves at different temperatures. The bottom insets show $M-M_{\text{fit}}(H)$ vs. temperature ($T$).

data. The insets at the lower left corners of Figure 4 display the detailed process for getting the $\mu_0H_{c1}$s at different measuring temperatures. Here we use the optimal superconducting CuIr$_2$Te$_{3.9}$I$_{0.1}$ as an example, as shown in Figure 4 (a). The applied field magnetization measurements $M(H)$ were performed at 2.4, 2.5, 2.6 and 2.8 K. The demagnetization effect should be considered to get accurate $Hc_1$ values. We can get the demagnetization factor ($N$) values from the formula $N = 4\pi\chi_V + 1$, where $\chi_V = dM/dH$ is the value of the linearly fitted slope (see the top inset in Figure 4 (a), the obtained $N$ value is around 0.48 to 0.57. The experimental data can be plotted using the formula $M_{fit} = e + fH$ at low magnetic fields, where $e$ is the intercept and $f$ denotes the slope of the linear fitting for the low magnetic field of $M(H)$ data. The bottom inset of Figure 4 (a) shows the curve for $M-M_{fit}$ ($H$) data versus the applied magnetic field ($H$). $\mu_0H_{c1}^*$ is estimated from 1 % $M$ at the field when it deviates below the fitted data ($M_{fit}$), which is the most common method.[12] The lower critical field $\mu_0H_{c1}(T)$ can be obtained with the consideration of the demagnetization factor ($N$) using the expression: $\mu_0 H_{c1}(T) = \mu_0 H_{c1}^*(T)/(1 - N)$ [46,47]. The $\mu_0H_{c1}(T)$ data can be further calculated from $\mu_0 H_{c1}(T) = \mu_0 H_{c1}(0)(1 - (T/T_c)^2$ . $\mu_0H_{c1}(0)$ values for $CuIr_2Te_{3.9}I_{0.1}$, $CuIr_2Te_{3.85}I_{0.15}$ and $CuIr_2Te_{3.2}I_{0.8}$ bulk are 0.024, 0.021 and 0.019 T, respectively, which are lower than that of the pristine compound $CuIr_2Te_4$ as summarized in Table 2.

We further examine the upper critical field ($\mu_0H_{c2}$) through temperature dependent measurements of the electrical resistivity, specific heat capacity $C(T, H)/T$ under various applied magnetic fields and the field-dependent magnetization $M(H, T)$ at various temperatures. The derived $\mu_0H_{c2}$ values of $CuIr_2Te_{3.9}I_{0.1}$, $CuIr_2Te_{3.85}I_{0.15}$ and $CuIr_2Te_{3.8}I_{0.2}$ as a function of the temperature are plotted in Figure 5 (a-c). The upper critical fields $\mu_0H_{c2}$s are determined through the temperature-dependent resistivity $\rho(T, H)$ using 10 %, 50 % and 90 % criteria of the normal-state resistivity values (see Figure 5 (g)). Upon applying a magnetic field, the superconducting transitions in both the $\rho(T)$ and the specific heat $C_p(T)/T$ data shift towards lower temperatures (see Figure 5 (d-f) and Figure 3 (b)). The $\mu_0H_{c2}$ values are calculated based on Werthamer-Helfand-Hohenburg (WHH) and Ginzberg-Landau (GL) theories. We use the simplified WHH equation as follows: $\mu_0H_{c2}(0) = -0.693T_c (dH_{c2}/dT)|_{T_c}$ for the dirty-limit superconductivity, [48-52] where $(dH_{c2}/dT_c)$ denotes the slope of $\mu_0H_{c2}(T)$ near $T_c$. The obtained values of $\mu_0H_{c2}(0)$ from WHH model for $CuIr_2Te_{4-x}I_x$ ($x$ = 0.1, 0.15 and 0.2) from the 50 % criteria are 0.188, 0.179 and 0.161 T, respectively. The obtained upper critical field values do not exceed the Pauli limiting field for the weak-coupling BCS superconductors $H^P = 1.86*T_c$ [53]. Hence, the values of $H^P$ can be estimated to be 5.543, 5.375 and 5.338 T, respectively. These values are larger than that of the undoped $CuIr_2Te_4$. Then, the Ginzburg-Landau coherence length ($\xi_{GL}(0)$) is calculated from this equation $H_{c2} = \phi_0/(2\pi\xi_{GL}^2)$ using the $H_{c2}(0)$ data of 50 % criteria of $\rho_N$ based on the WHH model, where $\phi_0$ is the flux quantum. The obtained values of $\xi_{GL}(0)$s for $CuIr_2Te_{3.9}I_{0.1}$, $CuIr_2Te_{3.85}I_{0.15}$ and $CuIr_2Te_{3.8}I_{0.2}$ are 41.9, 43.64 and 46.40 nm, respectively. Similar to the WHH Model, we calculated the upper critical field values by taking the criteria 10 %, 50 % and 90 % of $\rho_N$ with the following GL equation:[54] $\mu_0H_{c2}(T) = \mu_0H_{c2}(0)*[(1 - T/T_c)^2]/[(1 + T/T_c)^2]$ . The $\mu_0H_{c2}(0)$ values from 50 % $\rho_N$ criteria are calculated to be 0.232, 0.212, 0.209 T, respectively, which are all higher than that of the undoped $CuIr_2Te_4$. On the other hand, the $\mu_0H_{c2}$ values calculated by the WHH model are smaller compared with those corresponding values calculated by the GL model. Besides, it is worth mentioning that

10the $\mu_0H_{c2}(0)$ values obtained from $M(H, T)$, $C_p(T)/T$ are also lower than values estimated from $\rho(T, H)$.

We finally summarize the electronic phase diagram as a function of temperature and doping level $x$ for iodine-doped CuIr$_2$Te$_4$. As shown in Figure 6, a dome-like variation of superconducting critical temperature that is broad in composition can be observed: I substitution for Te can first enhance the superconducting transition temperature ($T_c$) and rises to 2.95 K at $x = 0.1$, followed by a decrease of $T_c$ and then disappearance when at $x \geq 1$. Nevertheless, the CDW-like transition was first immediately suppressed with a small amount doping content $x$ but reappeared in the doping range of $x > 0.9$. Though, the CDW signature in the resistivity becomes obvious while no superconductivity was observed for high $x$ lever ($0.9 \leq x \leq 1.0$). This phenomenon is different from the case of CuIr$_{2-x}$Ru$_x$Te$_4$, in which the CDW transition completely disappears in the whole Ru doping.

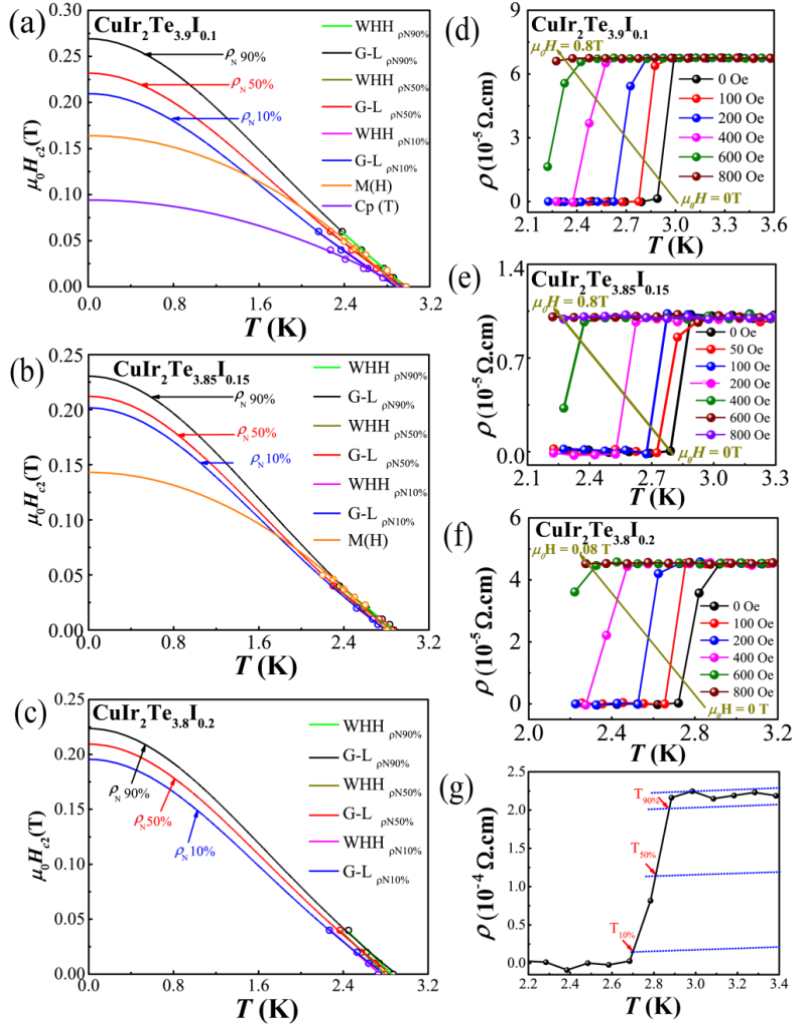

**Fig. 5. (a-c)** The temperature dependent upper critical fields extracted from the resistivity curves. The data are fitted by WHH and GL models for different criteria extracted from $\rho(T, H)$, $M(H, T)$ and $C_p(T, H)$. **(d-f)** The temperature dependence of the resistivity under different magnetic field $\rho(T, H)$ curves for CuIr$_2$Te$_{3.9}$I$_{0.1}$, CuIr$_2$Te$_{3.85}$I$_{0.15}$ and CuIr$_2$Te$_{3.8}$I$_{0.2}$, respectively. **(g)** Definition of the criteria $\rho_{N90\%}$, $\rho_{N50\%}$, and $\rho_{N10\%}$ of the superconducting transition.

Table 2. Comparison of superconducting parameters of telluride chalcogenide compounds.

| Compound | $CuIr_2Te_{3.9}I_{0.1}$ | $CuIr_2Te_{3.85}I_{0.15}$ | $CuIr_2Te_{3.8}I_{0.2}$ | $CuIr_2Te_{3.2}I_{0.8}$ | $CuIr_2Te_4$ [30] | $CuIr_{1.95}Ru_{0.05}Te_4$ [31] |
|---|---|---|---|---|---|---|
| $T_c$ (K) | 2.95 | 2.85 | 2.79 | 2.45 | 2.5 | 2.79 |
| $\gamma$ (mJ mol$^{-1}$ K$^{-2}$) | 12.97 | | | | 12.05 | 12.26 |
| $\beta$ (mJ mol$^{-1}$ K$^{-4}$) | 3.03 | | | | 1.97 | 1.87 |
| $\Theta_D$ (K) | 164.98 | | | | 190.3(1) | 193.6(2) |
| $\Delta C/\gamma T_c$ | 1.46 | | | | 1.50 | 1.51 |
| $\lambda_{ep}$ | 0.70 | | | | 0.63 | 0.65 |
| $N(E_F)$ (states/eV f.u) | 3.24 | | | | 3.10 | 3.15 |
| $\mu_0 H_{c1}(0)$ (T) | 0.024 | 0.021 | | 0.0188 | 0.028 | 0.098 |
| $\mu_0 H_{c2}(0)$ (T) ($\rho_N$ 50 % G-L theory) | 0.232 | 0.212 | 0.209 | | 0.145 | |
| $\mu_0 H_{c2}(0)$ (T) ($\rho_N$ 50% WHH theory) | 0.188 | 0.179 | 0.161 | | 0.12 | 0.247 |
| $-dH_{c2}/dT_c$ (T/K) | 0.092 | 0.091 | 0.083 | | 0.066 | 0.125 |
| $\mu_0 H^P$ (T) | 5.543 | 5.375 | 5.338 | | 4.65 | 5.24 |
| $\xi_{GL}$ (nm) ($\rho_N$ 50 % WHH theory) | 41.9 | 43.64 | 46.40 | | 52.8 | 36.3 |

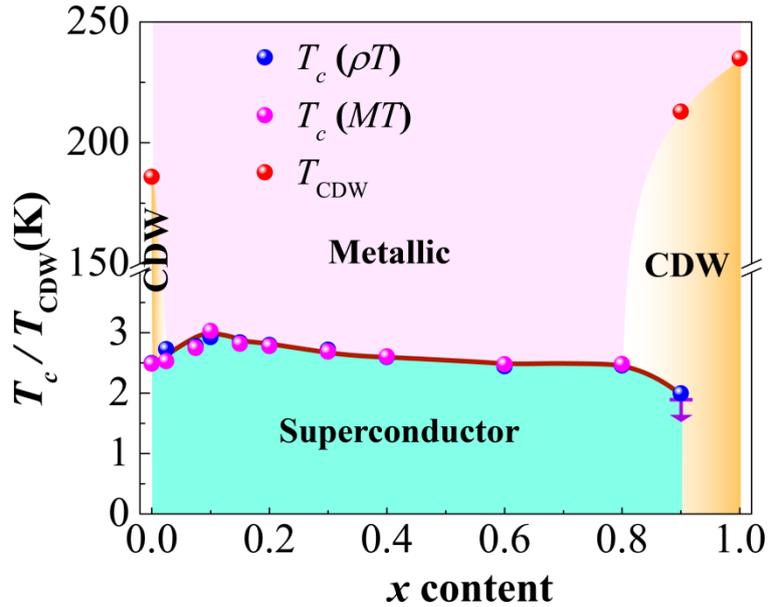

**Fig. 6.** The electronic phase diagram for $CuIr_2Te_{4-x}I_x$ (0.0 ≤ $x$ ≤ 1.0).

In summary, we have investigated the influence of iodine doping on the superconducting and CDW properties of $CuIr_2Te_{4-x}I_x$ compound. The transport, magnetic and specific heat measurements reveal that iodine doping can enhance the superconducting transition while the CDW transition was first suppressed at low $x$ but reoccurred at high iodine doping region. The optimal superconducting composition is $CuIr_2Te_{3.9}I_{0.1}$, which is confirmed to be a type-II superconductor with $T_c$ ~ 2.95 K. The upper critical field for this optimal superconducting $CuIr_2Te_{3.9}I_{0.1}$ is higher than that of the pristine compound; while its lower critical field is lower than that of the undoped compound. A "dome-like" electronic phase diagram has been established, in which the CDW transition has been suppressed promptly at $x = 0.025$ and $T_c$ increased up to 2.95 K ($x = 0.1$), followed by a slight decrease to 2.43 K ($x = 0.8$); while the CDW showed up again at high $x \sim 0.9$ to 1.0. However, the reason why the CDW state can be suppressed by very low iodine doping but become obviously at the high doping content $x$ is still unknown. An intensive study to understand the present experimental findings is necessary.

# Supporting Information

# Superconductivity and Charge Density Wave in Iodine-doped CuIr$_2$Te$_4$

Mebrouka Boubeche[1†], Jia Yu(于佳)[2†], Chushan Li(李楚善)[2], Huichao Wang(王慧超)[2], Lingyong Zeng(曾令勇)[1], Yiyi He(何溢懿)[1], Xiaopeng Wang(王晓鹏)[1], Wanzhen Su(苏婉珍)[1], Meng Wang(王猛)[2], Dao-Xin Yao(姚道新)[2], Zhijun Wang(王志俊)[3,4] and Huixia Luo(罗惠霞)[1**]

[1]School of Materials Science and Engineering, State Key Laboratory of Optoelectronic Materials and Technologies, Key Lab of Polymer Composite & Functional Materials, Sun Yat-Sen University, No. 135, Xingang Xi Road, Guangzhou, 510275, P. R. China
[2]School of Physics, Sun Yat-Sen University, No. 135, Xingang Xi Road, Guangzhou, 510275, P. R. China
[3]Beijing National Laboratory for Condensed Matter Physics, and Institute of Physics, Chinese Academy of Sciences, Beijing 100190, China
[4]University of Chinese Academy of Sciences, Beijing 100049, China
† These authors contributed equally to this work; email: boubeche@mail.sysu.edu.cn; yujia7@mail.sysu.edu.cn
** Corresponding author Email: luohx7@mail.sysu.edu.cn; Tel: (+0086) 13802768250

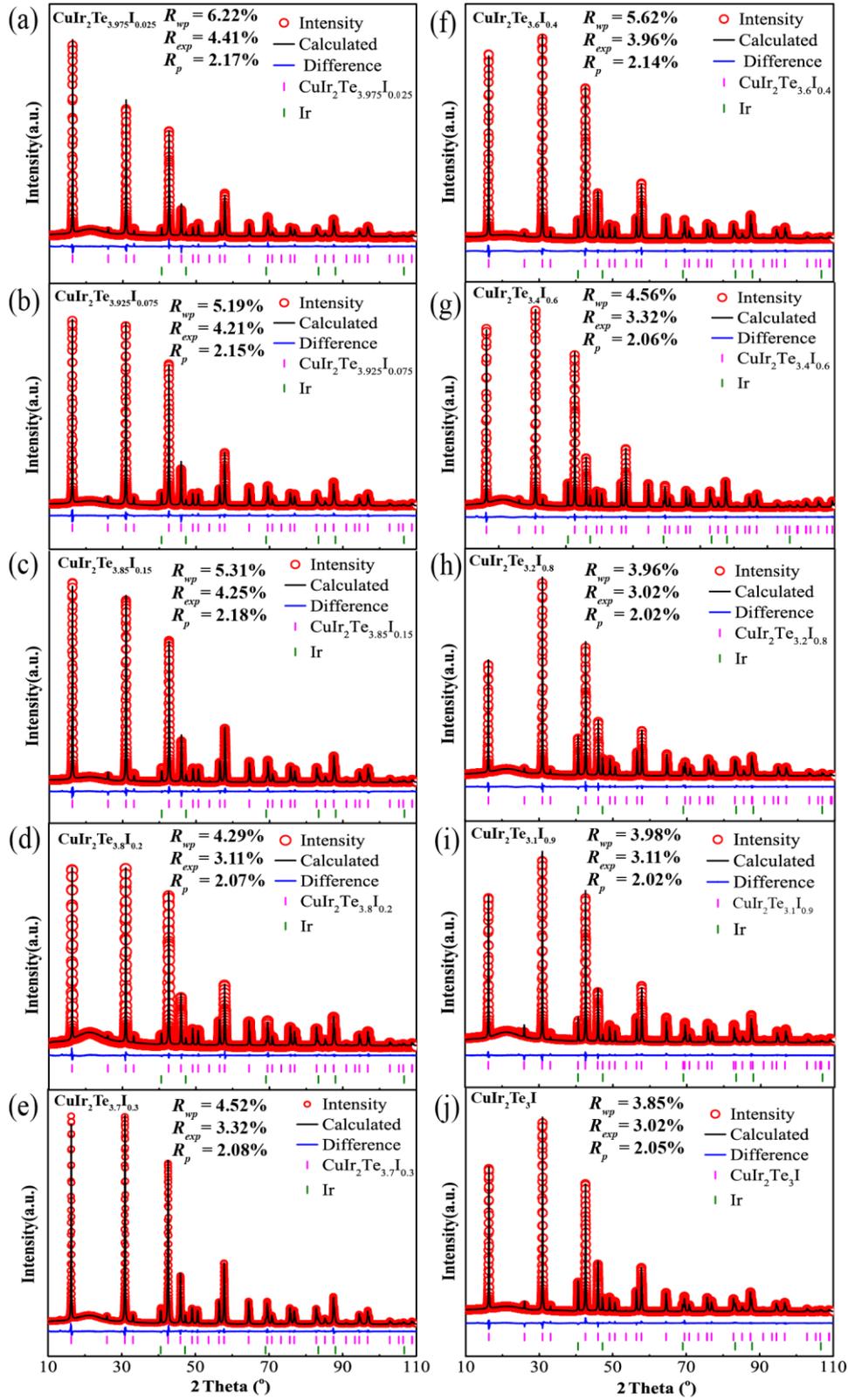

**Fig. S1.** The refinement graphs of CuIr$_2$T$_{4-x}$I$_x$ polycrystalline samples.

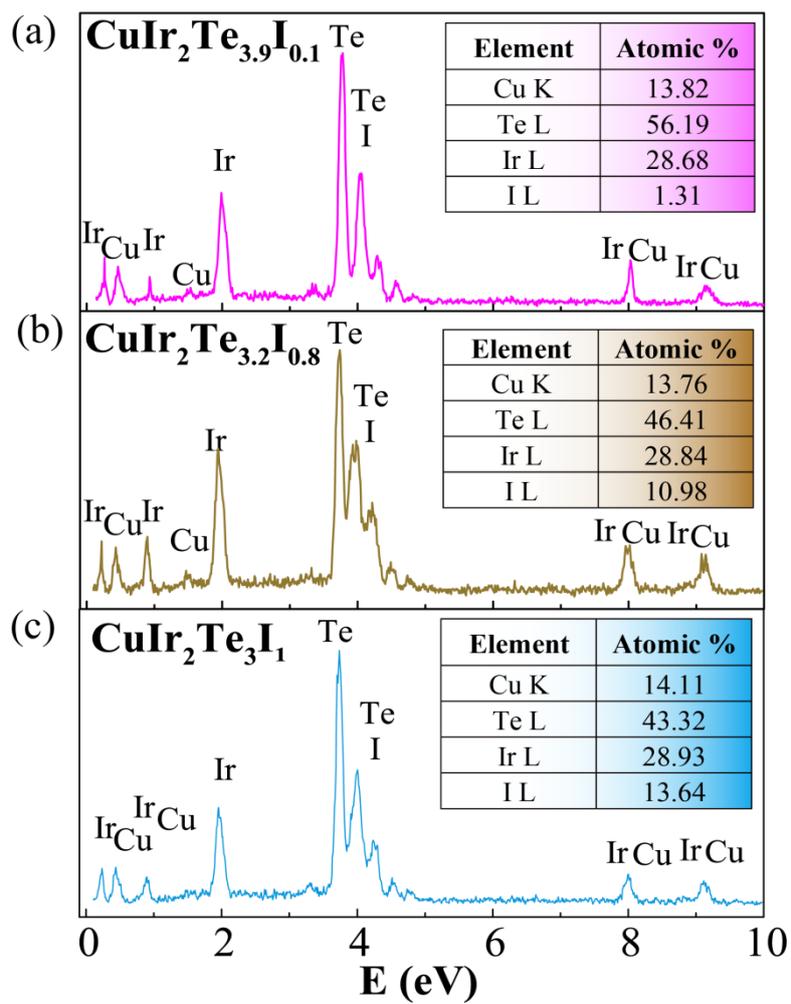

**Fig. S2.** EDXS spectrum of $CuIr_2Te_{4-x}I_x$, (a) $x = 0.1$, (b) $x = 0.8$, (d) $x = 1.0$, insets are the corresponding atomic ratios of the elements.

**Tab. S1.** The element ratios of $CuIr_2Te_{4-x}I_x$ from EDXS results.

| Element ratio<br>Sample | Cu | Ir | Te | I |
|---|---|---|---|---|
| $CuIr_2Te_4$ [30] | 0.97 | 1.96 | 3.93 | 0 |
| $CuIr_2Te_{3.9}I_{0.1}$ | 0.98 | 1.97 | 3.86 | 0.09 |
| $CuIr_2Te_{3.2}I_{0.8}$ | 0.97 | 1.96 | 3.17 | 0.75 |
| $CuIr_2Te_3I_1$ | 0.95 | 1.95 | 2.92 | 0.92 |

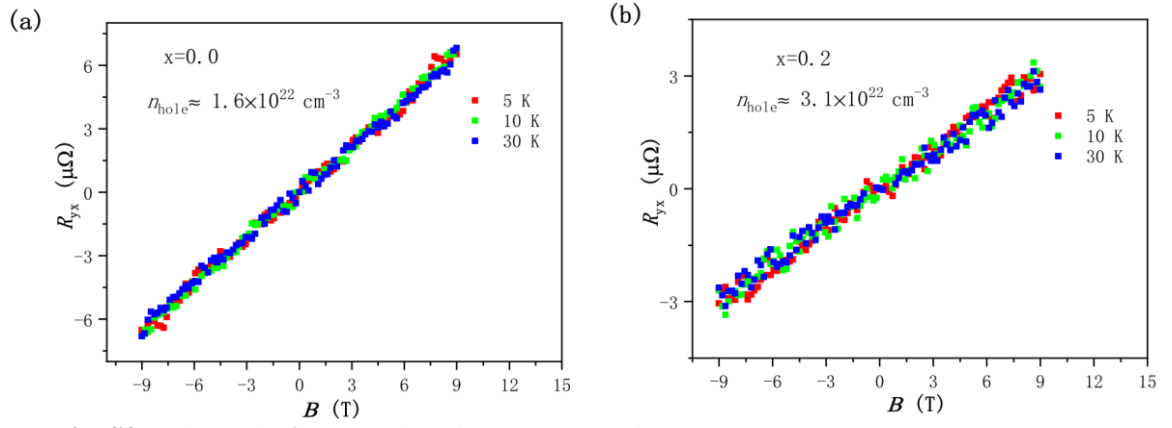

**Fig. S3.** Hall results for (a) undoped $CuIr_2Te_4$ and (b) $CuIr_2Te_{3.8}Te_{0.2}$ at low temperatures.